\documentclass[twocolumn]{aastex62}

\usepackage{amsmath} 
\usepackage{natbib}
\usepackage{url}
\graphicspath{{./}{figures/}}

\def\I{\,\textsc{i}}

\begin{document}

\title{The Type II-P Supernova 2019mhm and Constraints on Its Progenitor System}
\correspondingauthor{Jason Vazquez}
\email{jasonv3@illinois.edu}

\author{J. Vazquez}
\affil{Department of Astronomy, University of Illinois at Urbana-Champaign, Urbana, IL 61801, USA}

\author{C.\ D.\ Kilpatrick}
\affiliation{Center for Interdisciplinary Exploration and Research in Astrophysics (CIERA) and Department of Physics and Astronomy, Northwestern University, 2145 Sheridan Road, Evanston, IL 60201, USA}

% Adding other authors who need to be on the paper
\author{G. Dimitriadis}
\affiliation{School of Physics, Trinity College Dublin, The University of Dublin, Dublin 2, Ireland}
% dimitrig@tcd.ie
\author{R. J. Foley}
\affiliation{Department of Astronomy and Astrophysics, University of California, Santa Cruz, CA 95064, USA}
% foley@ucsc.edu
\author{A. L. Piro}
\affiliation{The Observatories of the Carnegie Institution for Science, 813 Santa Barbara St., Pasadena, CA 91101, USA}
% piro@carnegiescience.edu
\author{A. Rest}
\affiliation{Space Telescope Science Institute, 3700 San Martin Drive, Baltimore, MD 21218}
\affiliation{Department of Physics and Astronomy, Johns Hopkins University, 3400 North Charles Street, Baltimore, MD 21218, USA}
% arest@stsci.edu
\author{C. Rojas-Bravo}
% crojasbr@ucsc.edu
\affiliation{Department of Astronomy and Astrophysics, University of California, Santa Cruz, CA 95064, USA}

\begin{abstract}

We present pre- and post-explosion observations of the Type II-P supernova (SN~II-P) 2019mhm located in NGC~6753. Based on optical spectroscopy and photometry, we show that SN\,2019mhm exhibits broad lines of hydrogen with a velocity of $-8500\pm200$~km~s$^{-1}$ and a $111\pm2$~day extended plateau in its luminosity, typical of the Type II-P subclass. We also fit its late-time bolometric light curve and infer that it initially produced a ${}^{56}$Ni mass of $1.3 \times 10^{-2}$~M$_{\odot} \pm 5.5 \times 10^{-4}$~$M_\odot$. 
Using imaging from the Wide Field Planetary Camera 2 on the {\it Hubble Space Telescope} obtained 19~years before explosion, we aligned to a post-explosion Wide Field Camera 3 image and demonstrate that there is no detected counterpart to the SN to a limit of $>$24.53~mag in F814W, corresponding to an absolute magnitude limit of $M_{\rm F814W} < -7.7$~mag. Comparing to massive-star evolutionary tracks, we determine that the progenitor star had a maximum zero-age main sequence mass $<$17.5~M$_{\odot}$, consistent with other SN~II-P progenitor stars. SN\,2019mhm can be added to the growing population of SNe~II-P with both direct constraints on the brightness of their progenitor stars and well-observed SN properties.

\end{abstract}

\keywords{stars: evolution --- supernovae: general --- supernovae: individual (SN~2019mhm)}

\section{Introduction} 
\label{sec:intro}

The core of a star of at least 8~$M_{\odot}$ will catastrophically collapse at the end of its life typically leading to a supernova (SN) \citep{Burrows95}. Luminous supernovae (SNe) observed with hydrogen in their spectra are classified as Type II, with further sub-classifications called Type II-P, Type II-L, and Type IIb based on light curve and spectroscopic properties \citep{Barbon79, Filippenko97, Arcavi17}. The most common Type II-P SNe quickly reach maximum brightness and then their luminosities tend to stay relatively constant in optical bands, resulting in a ``plateau'' in their light curves due to the recombination of hydrogen in their massive outer envelopes \citep{Falk77}. Nickel starts to decay into cobalt, and the SN then fades significantly to luminosities powered by radioactive $^{56}$Co. Thus Type II-P SNe require progenitor stars with massive, extended hydrogen envelopes, which has long been taken as an indication that they arise from red supergiant (RSG) stars \citep{Arnett87, Woosley87, Falk73}.

Observations of RSG stars in the Milky Way and nearby galaxies suggest they range from ZAMS masses of 8--30~$M_{\odot}$ \citep{Davies18a, Davies18b, Davies20}. Dozens of RSGs in nearby galaxies have been directly detected and confirmed as pre-explosion counterparts to Type II-P SNe \citep{Smartt04, MaundSmartt09, Maund09, Fraser10, Fraser11, Fraser14, Crockett11, Elias11,Maund13, Maund14, Tom13, Kochanek17}. All of these RSG progenitor stars have zero-age main sequence (ZAMS) masses of $<$17~M$_{\odot}$ with fewer than expected high-mass counterparts to any SN type found in pre-explosion imaging, despite expecting more massive stars given a Salpeter Initial Mass Function \citep[the ``red supergiant problem'';][]{Smartt09,Smartt15}. Whether the SN progenitor-star mass function is actually cut off at or around 17-20~$M_{\odot}$ or there is some systematic or observational effect associated with this finding is unclear. For example, one proposed solution to the RSG problem is that some higher mass stars implode as ``failed SNe,'' becoming black holes producing a low-luminosity red transient with lower ejecta mass than a typical SNe \citep{Woosley86, Oconnnor11, Adams17, Piro13}. Other proposed solutions include the possibility that higher-mass stars shed their hydrogen envelopes and explode as other types of SNe \citep{Smith11,Beasor16}, reducing the contribution of high-mass RSGs to the SNe progenitor population. 

A common assumption in addressing the red supergiant problem is that the line-of-sight extinction to the observed progenitor star is the same as the extinction inferred to the resulting SN. For example, the extinction to the progenitor star could be larger if the resulting SN destroys dust confined close to the star, or it could be smaller if the star produces dust between the time of the pre-explosion imaging and onset of the SN. \cite{Fraser12} and \cite{VanDyk12a} both found significant reddening for the progenitor of SN~2012aw, and concluded it was from circumstellar dust that was eventually destroyed after the explosion. However, \cite{Walmswell12} calculated additional extinction due to dust production from SN winds, but this hypothetical extinction was insufficient to explain the apparently missing high-mass stars in the red supergiant problem. It should be noted that \cite{Kochanek12} demonstrated the physical inconsistencies when treating extinction of circumstellar material with interstellar medium material due to scattering of off-axis light in compact dust shells, potentially exaggerating the effect of extinction from this circumstellar material. Additional, well-observed examples are essential to disentangle the diversity of effects this may have on SNe and their massive progenitor stars.

There is still ambiguity on what the exact distribution of progenitors is and how high the upper mass limit can go, leaving a need for additional Type II progenitor detections and upper limits to constrain the intrinsic distribution. In this paper we present SN\,2019mhm, a Type II-P SN discovered in NGC~6753 by Backyard Observatory Supernova Search (BOSS) transient survey 2019 Aug 2 13:13:21 \citep[UT;][]{Marples19}, located at $\alpha$ $ = 19^{\rm h}11^{\rm m}24^{\rm s}.06$, $\delta$ = $-57^{\circ}03^{\prime}18^{\prime\prime}.00$ (J2000), with an apparent magnitude $m_{C} = 16.6$~Vega mag using a clear filter. We present optical observations of SN\,2019mhm and constraints on the luminosity of its progenitor star using pre- and post-explosion {\it Hubble Space Telescope} imaging. The distance assumed for NGC~6753 is $23.4 \pm 5.3$~Mpc via the fundamental plane method \citep{Springob14}, along with an inferred distance modulus of $DM = 31.85$~mag. We adopt a redshift of $z=0.01057$ \citep{Wong06}. 

\section{OBSERVATIONS}

\subsection{{\it Hubble Space Telescope} Imaging}

\begin{figure*}
\begin{center}
\includegraphics[width=\textwidth]{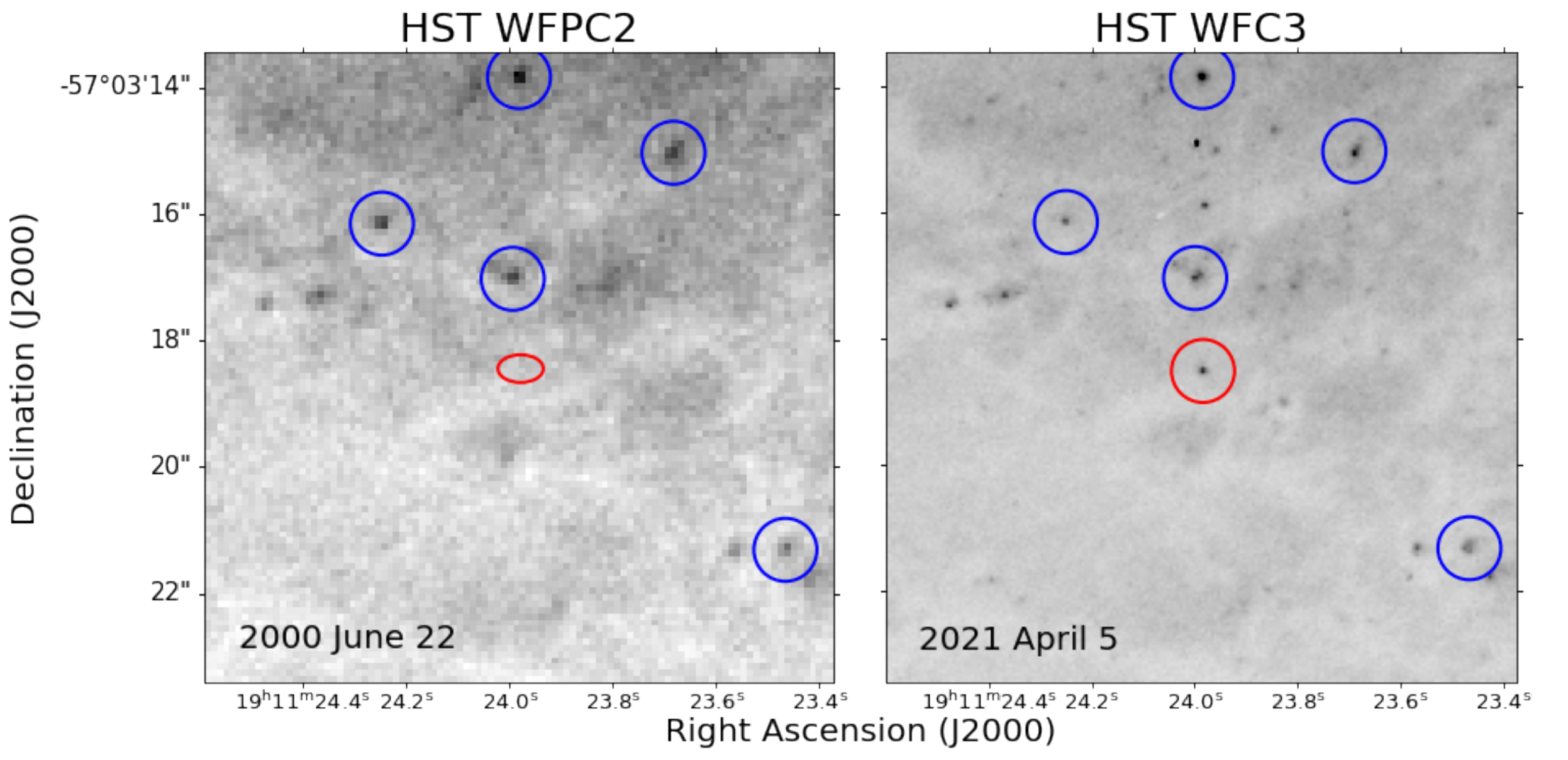} 
\caption{WFPC2 and WFC3 imaging of a 10.0$^{\prime\prime}\times$ 10.0$^{\prime\prime}$ region centered around where we would expect to see the pre-explosion counterpart to SN~2019mhm, taken 2000 June 22 (pre-explosion,left) and WFC3 imaging of a region around SN~2019mhm taken 2021 April 05 (post-explosion, right). Both images are using the HST F814W filter. The location of SN~2019mhm is outlined by a red ellipse at $20\sigma$ in the left panel, and circled in red in the right panel. 5 additional point sources are circled in blue. \label{fig:1}}
\end{center}
\end{figure*}

We obtained imaging of NGC~6753 from the {\it Hubble Space Telescope} ({\it HST}) Mikulski Archive for Space Telescopes covering the site of SN\,2019mhm. The data were taken on 2000 June 22 20:26:14 using the Wide Field and Planetary Camera 2 (WFPC2) and on 2021 April 05 20:07:52 UT using the Wide Field Camera 3 (WFC3). WFPC2 observed for 320 seconds in the F814W filter and 1900 seconds in the F300W filter (Proposal ID 8645, PI Windhorst). We observed with WFC3 for 780 seconds in F814W and 710 seconds in F300X (Proposal ID 16239, PI Foley). The specific observations analyzed can be accessed via \dataset[10.17909/ggct-aw51]{https://doi.org/10.17909/ggct-aw51}. We then ran the files through the {\it HST} imaging and photometry pipeline {\tt hst123} \citep{Kilpatrick21} to align and construct drizzled frames for each unique filter and epoch. There were 148 common sources found in F814W. Using those combined images we further aligned each image using {\tt TweakReg} \citep{tweakreg}, which uses a dynamic signal-to-noise threshold and a Gaussian convolution kernel with a 2.5 or 3.5 pixel full width at half maximum depending on the instrument to pick bright, isolated point sources. This resulted in a frame-to-frame alignment precision of 0.047 arcsec in right ascension and 0.036 arcsec in declination. To estimate the systematic uncertainty on our alignment procedure, we repeated the process using the same set of astrometric standards and splitting the sample randomly in half. Using half the stars to calculate a new alignment solution from WFPC2 to WFC3, we calculated the root-mean square offset in right ascension and declination between the positions of the other half of the stars in the WFC3 image and their updated positions in the aligned WFPC2 image. Repeating this process 100 times, we calculate the average root-mean square offset across each trial, which we consider to be the total systematic uncertainty in alignment. Our final alignment uncertainty is $0.36~({\rm stat.}) + 0.31~({\rm sys.})$ WFPC2 pixels in right ascension and $0.22~({\rm stat.}) + 0.29~({\rm sys.})$ WFPC2 pixels in declination. We obtained photometry of all sources near the site of SN\,2019mhm using {\tt dolphot} \citep{dolphot}. The pre- and post-explosion F814W images centered on the site of SN\,2019mhm are shown in \autoref{fig:1}.

\subsection{Imaging of SN\,2019mhm}

We observed SN\,2019mhm in $uBV\!gri$ bands with the Swope 1-m optical telescope at Las Campanas Observatory, Chile from 2 August 2019 to 14 March 2020. All data were calibrated using bias and flat-field frames observed in the same instrumental configuration as described in \citet{Kilpatrick18b} and using the {\tt photpipe} imaging and photometry package \citep{Rest05}. We used {\tt photpipe} to perform amplifier crosstalk corrections, masking, astrometric calibration, photometry using a custom version of {\tt DoPhot} \citep{Schechter93}, and photometric calibration using SkyMapper photometric standards \citep{skymapper} observed in the same field as SN\,2019mhm. Finally, using deep template imaging obtained on 9--10 October 2021, we performed image subtraction using {\tt hotpants} \citep{hotpants} and forced photometry on the position of SN\,2019mhm using a custom version of {\tt DoPhot}.

\begin{figure}
\begin{center}
\includegraphics[width=0.49\textwidth]{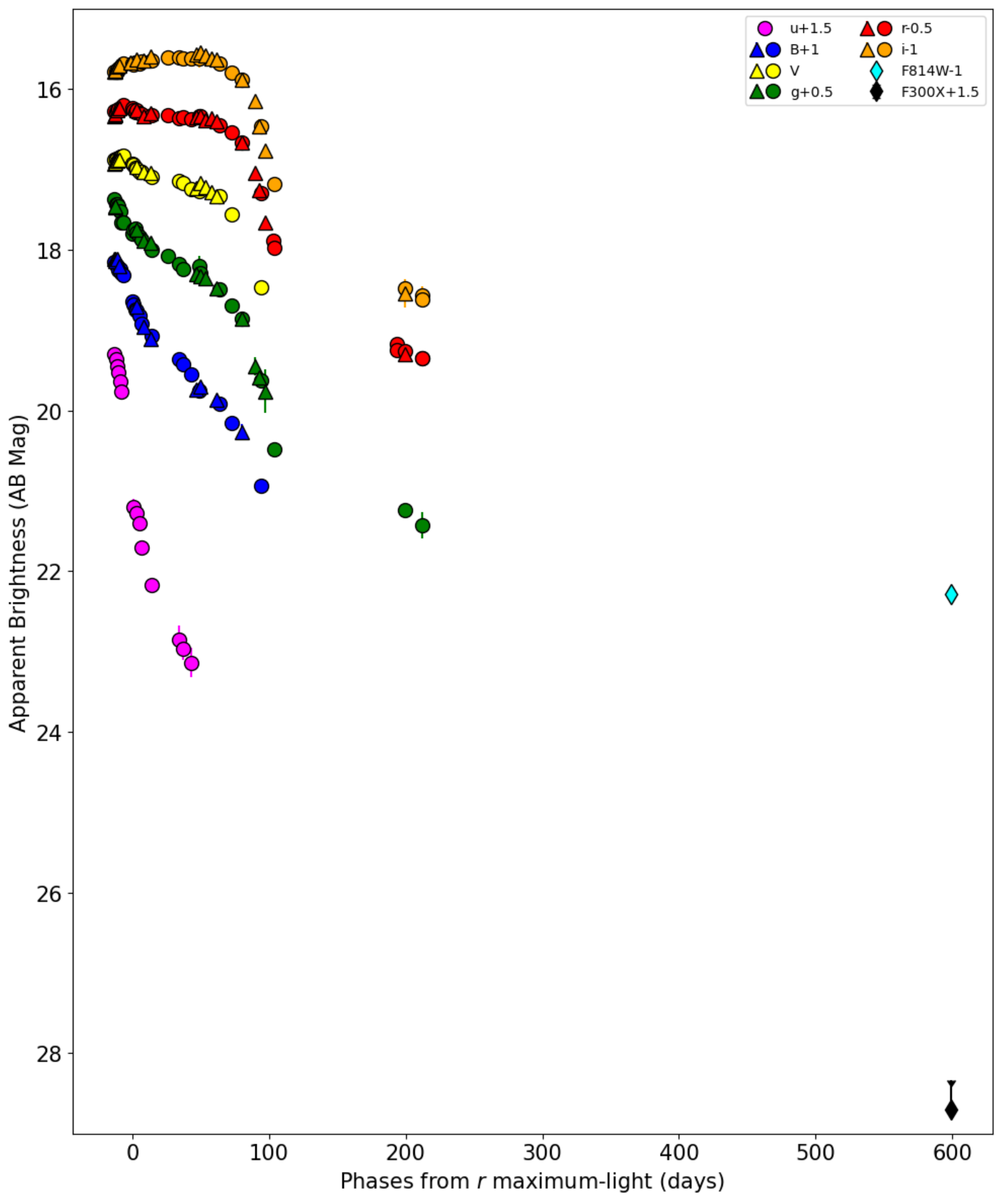}
\caption{Light curves of SN\,2019mhm from {\it HST} (diamonds), LCO (triangles), and Swope (circles) photometry of SN~2019mhm covering $uBVgri$, F814W, and F300X. Dates are reported with respect to maximum-light in the $r$ band.}\label{fig:lc}
\end{center}
\end{figure}

SN\,2019mhm was also observed in $BVg^{\prime}r^{\prime}i^{\prime}$ bands by the Sinistro imagers on the Las Cumbres Observatory (LCO) 1-m telescope network from 2019 August 2 to 2020 March 2. Starting from the pixel-level calibrated imaging provided by the LCO {\tt BANZAI} pipeline \citep{BANZAI}, we performed all masking, astrometric, and photometric calibration following identical methods to those described above for the Swope imaging. We then used the final epoch of $BVg^{\prime}r^{\prime}i^{\prime}$ LCO imaging obtained on 2020 June 18 as templates to perform image subtraction with {\tt hotpants} and obtained forced photometry with {\tt DoPhot}. The final forced-photometry of SN\,2019mhm from our Swope and LCO difference images is shown in \autoref{tab:cand-photometry} and \autoref{fig:lc}.

\subsection{Spectroscopy of SN\,2019mhm}

\begin{figure}
\begin{center}
\includegraphics[width=0.49\textwidth]{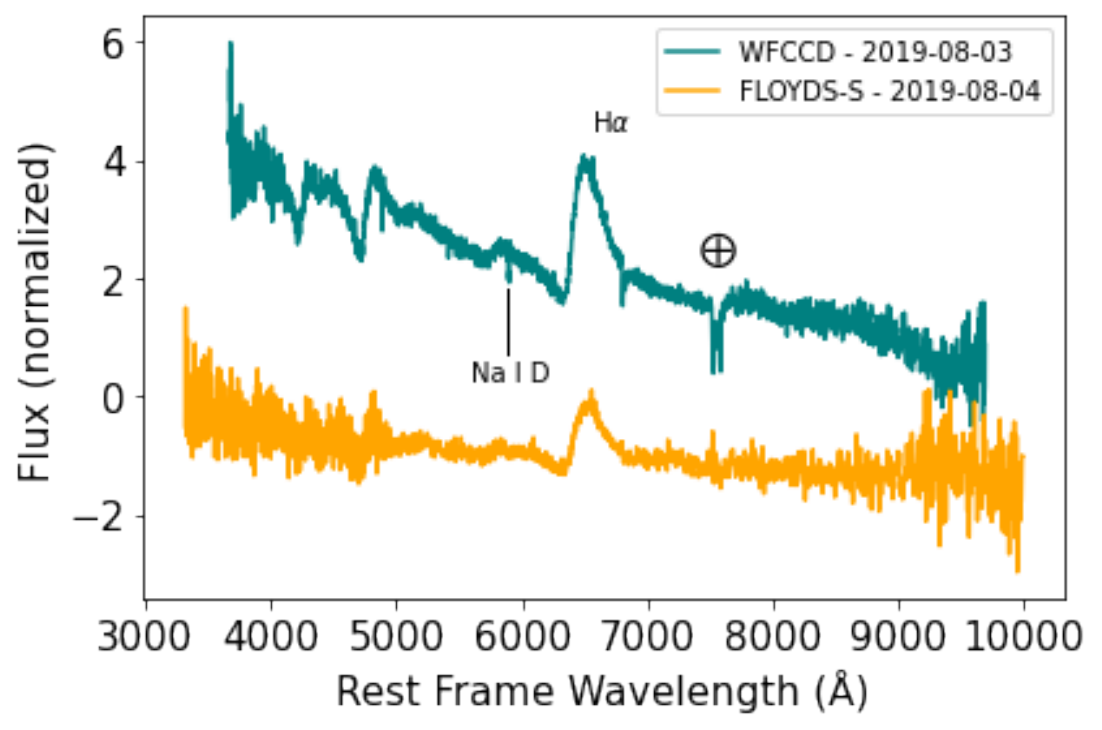}
\caption{Spectra of SN~2019mhm in WFCCD and FLOYDS-S. The H-$\alpha$ line, Na I D line, and Telluric feature are highlighted.}
\label{fig:spectra}
\end{center}
\end{figure}

We obtained the classification spectrum of SN\,2019mhm for our spectral analysis \citep[as described in][]{Chen19}. These data were obtained at Las Campanas Observatory with the du Pont 2.5m telescope and Wide field reimaging CCD (WFCCD) on 3 August 2019. All data were processed following standard procedures in {\tt IRAF}\footnote{IRAF is distributed by the National Optical Astronomy Observatory, which is operated by the Association of Universities for Research in Astronomy (AURA) under a cooperative agreement with the National Science Foundation.}, including flattening, aperture extraction, wavelength calibration, and flux calibration. In addition, we analyzed spectroscopy of SN\,2019mhm obtained through the Las Cumbres Observatory (LCO) network at Siding Spring Observatory, Australia on the Faulkes 2m telescope and FLOYDS spectrograph from 4 August 2019. We reduced these data with the LCO FLOYDS pipeline \citep{valenti_floyds_2013} \footnote{\url{https://github.com/LCOGT/floyds_pipeline}} using arc and flat-field frames obtained on the same night and in the same instrumental configuration. The final spectra are shown in \autoref{fig:spectra}.

\section{Methods and Results}\label{sec:methods}

\subsection{Spectroscopic and Extinction Properties of SN\,2019mhm}

The spectra of SN\,2019mhm exhibit broad lines of H$\alpha$ and H$\beta$, which combined with its peak magnitude and light-curve evolution highlight its nature as a Type II-P SN. For the Milky Way, we use the reddening value of $E(B-V)=0.06$~mag from \cite{Schlafly11}. Below we assume that the total-to-selective extinction in NGC~6753 is $R_{V}=A_{V}/E(B-V)=3.1$ \citep{Cardelli89},  similar to the Milky Way.

\begin{figure}
\begin{center}
\includegraphics[width=0.49\textwidth]{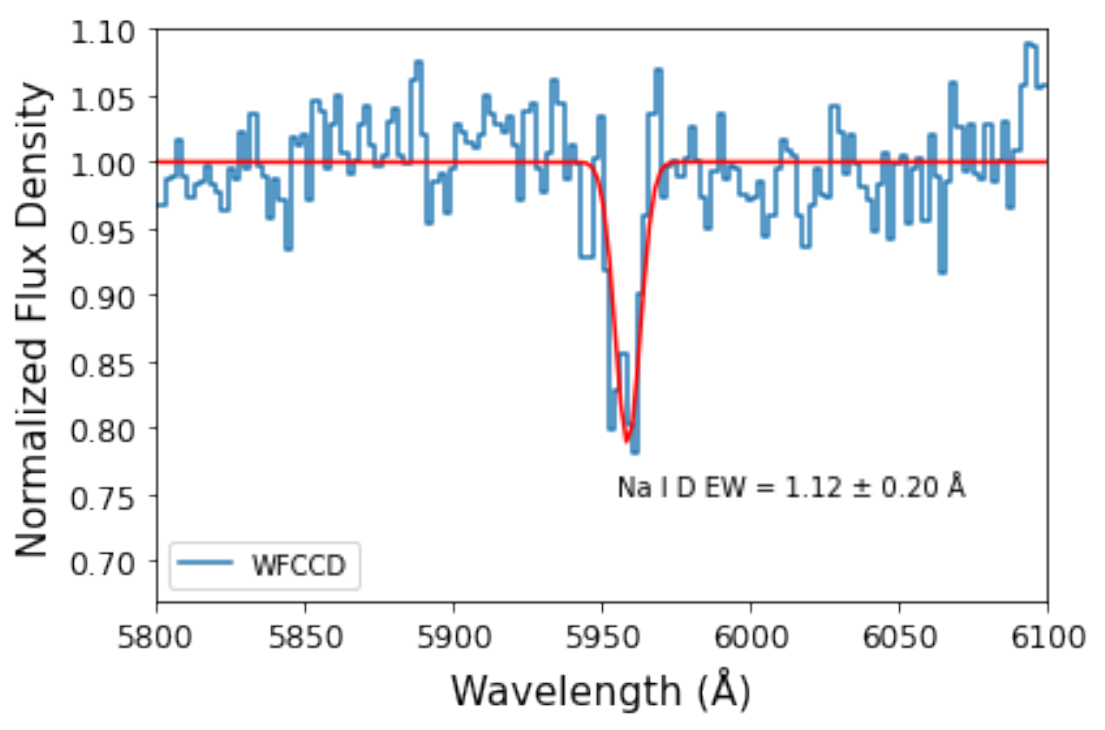}
\caption{Na I D doublet in WFCCD spectrum of 2019mhm. The equivalent width of the doublet is $1.12 \pm 0.20$~\AA.}
\label{fig:Na}
\end{center}
\end{figure}

It is possible to determine the internal extinction from the EW of the Na ID absorption line at the host galaxy redshift. To infer the Na\I~D equivalent width, we first removed the recessional velocity of NGC~6753 and normalized the continuum emission in the WFCCD spectrum. We masked the wavelength range around the Na\I~D doublet centroid (near $\lambda\lambda$5889.95, 5895.92). We then masked the continuum around the centroid, and we fit a Gaussian profile to the Na\I~D absorption profile. The resulting equivalent width was $1.12 \pm 0.20$~\AA, as shown in \autoref{fig:Na}. To convert this quantity to a reddening value, we used the relation between Na\I~D column density and extinction presented in \cite{Phillips13}. We adopt a host-galaxy reddening value to be $E(B-V)=0.180 \substack{+0.036\\-0.029}$~mag assuming $R_{V}=3.1$ for the following analyses. 

We performed a similar analysis to calculate the velocity of the much broader H$\alpha$ line. Here we masked the wavelength range around $\lambda$ = 6563~\AA. We calculate the velocity to be $-8500\pm200$~km~s$^{-1}$ at the absorption minimum. This velocity is consistent with those measured from other Type II SNe around or slightly before maximum light, for example well-observed events such as SNe\,1986L, 1999em ($9820$~km~s$^{-1}$), and 2012aw \citep[$12800$~km~s$^{-1}$;][]{Bose13} used to infer distances via the expanding photosphere method \citep{Schmidt94,Leonard02,Bose14}. 

\subsection{The Bolometric Light Curve of SN\,2019mhm}

Our multi-band light curves of SN\,2019mhm are shown in \autoref{fig:lc}, which track the evolution of this event from 12 hours to 612 days from discovery. We ran all light curves through {\tt superbol} \citep{superbol} to integrate the time-varying spectral energy distribution (SED) of SN\,2019mhm and obtain a bolometric light curve by integrating the black body fits of the SEDs over all wavelengths. We determined a bolometric luminosity at each $i$-band epoch, and performed a linear interpolation on the other light curves to that epoch to construct the full SED. We used a distance modulus value of $31.85$~mag and a reddening value of $E(B-V) = 0.24$~mag as mentioned above when running the photometry through {\tt superbol}, and we did not impose any UV suppression as our observations do not extend blueward of $u$-band. 

The final epoch for SN\,2019mhm was 612.845 days, corresponding to a rest-frame epoch of 606 days. We searched the Weizmann Interactive Supernova Data Repository \citep[WISseREP;][]{wiserep} for spectra of other SNe~IIP
with rest-frame epochs between 525 and 675 days, and found a single spectrum for the Type II-P SN\,2004dj at 665 rest-frame days from explosion. From this spectrum, we used {\tt pysynphot} \citep{pysynphot} to calculate a pseudo-bolometric magnitude and in-band magnitude for F814W and thus a pseudo-bolometric correction for that band, which we estimate to be $BC_{{\rm F814W}} = 0.35$~mag. This value is in agreement within the errors of our {\tt superbol} analysis assuming no color evolution from the previous epoch, which results in $BC_{{\rm F814W}}=-0.2\pm0.8$~mag.  We adopt $BC_{{\rm F814W}}=0.35$~mag for the analysis below.

\begin{figure} 
\begin{center} 
\includegraphics[width = 0.49\textwidth]{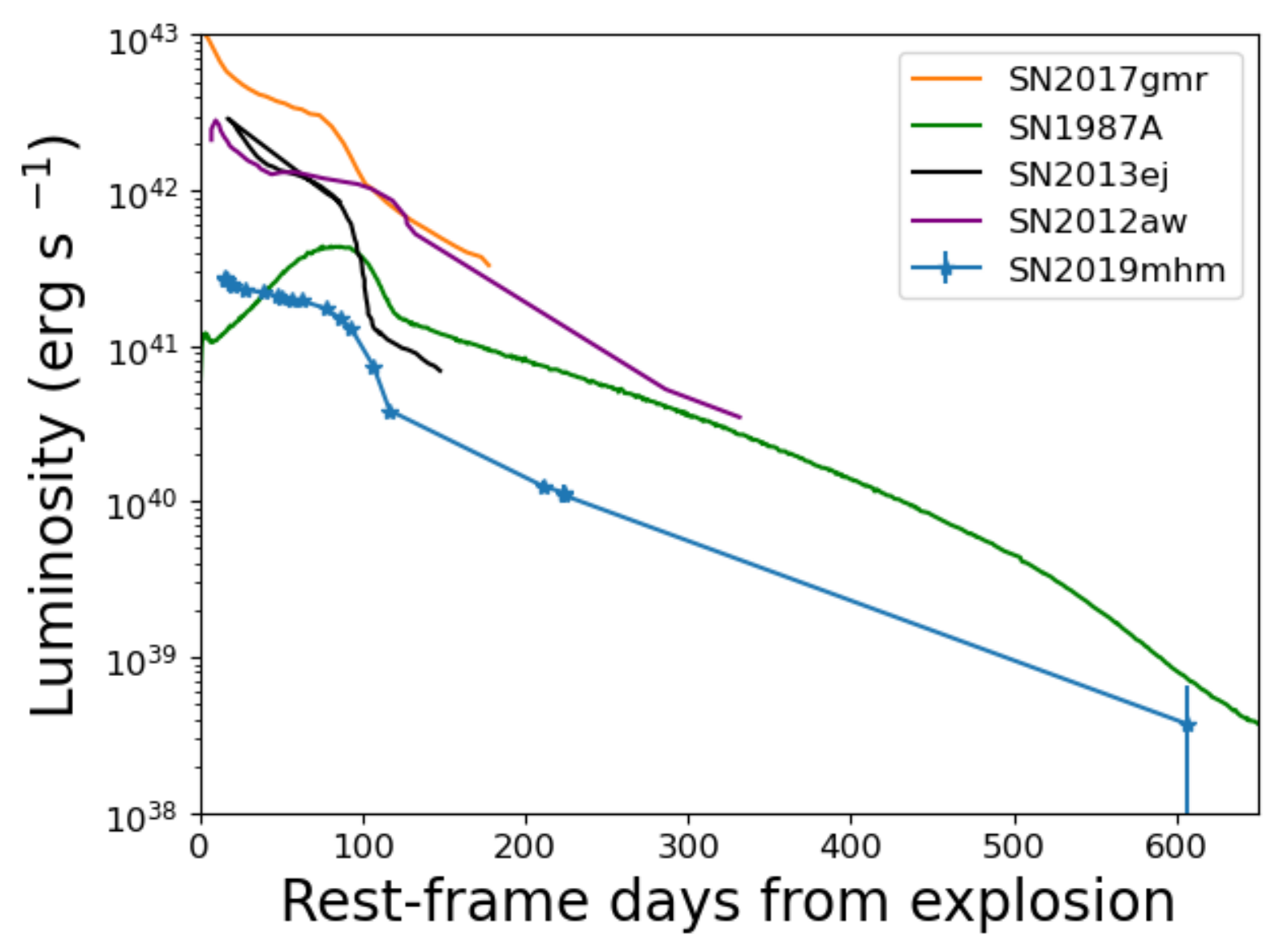}
\caption{Bolometric light curves of Type II-P SNe.}
\label{fig:LCs}
\end{center}
\end{figure}

We then plotted the bolometric light curve with respect to time from explosion for SN\,2019mhm using the  assuming the time of explosion was the discovery date in MJD, plotted in \autoref{fig:LCs}. We used the {\tt emcee} package \citep{MCMC} to infer the plateau duration using equation 1 from \cite{Valenti16}.The decline rate of SN\,2019mhm closely tracks expectations for a ${}^{56}$Co-powered light curve \citep[with a $\approx0.012$~mag~day$^{-1}$ decline as in][]{Valenti16}. This is at both $200$~days from explosion when deep ground-based observations were still possible and out to 610~days from explosion as observed by {\it HST}. We can therefore scale the bolometric light curve of SN\,2019mhm to other well-observed SNe such as SN\,1987A \citep[e.g.,][]{Spiro14} to infer the total ${}^{56}$Ni mass produced in the event. Using equation 3 of \cite{Spiro14} and interpolating the bolometric luminosities of SN\,2019mhm and SN\,1987A at 200 days from explosion, we derive a ${}^{56}$Ni mass of $1.3 \times 10^{-2}$~$M_\odot \pm 5.5 \times 10^{-4}$~$M_\odot$. 

We compare SN\,2019mhm's bolometric light curve to those of other Type II SNe in \autoref{fig:LCs}, including SN\,2017gmr \citep{17gmr}, SN\,1987A \citep{87a1,Suntzeff90, Hamuy90}, SN\,2013ej \citep{13ej}, and SN\,2012aw \citep{12aw}. SN\, 2019mhm is still brighter and with a higher ${}^{56}$ Ni mass than the objects belonging to the family of 'faint' or 'low-luminosity' Type II-P SNe \citep{Spiro14}. 

\subsection{Constraints on a pre-explosion counterpart to SN\,2019mhm}

 We identified SN\,2019mhm in a single post-explosion WFC3 F814W image, and consider a counterpart to be a source within a 3$\sigma$ radius of SN\,2019mhm's location. We did not find any counterpart within 3$\sigma$ alignment precision of that location in pre-explosion WFPC2 F814W imaging (\autoref{fig:1}). There are seven point-like sources within a 5\arcsec\ radius of the explosion site of WFPC2 imaging, the closest of which is approximately 1.1\arcsec\ (41$\sigma$ astrometric uncertainty) away and detected at $m_{\rm F814W}=22.6$~mag. We therefore consider there to be no pre-explosion counterpart to SN\,2019mhm in the WFPC2 imaging, and below we infer the upper limit on the flux to any counterpart in the pre-explosion data.

\begin{figure}
\begin{center}
\includegraphics[width = 0.49 \textwidth ]{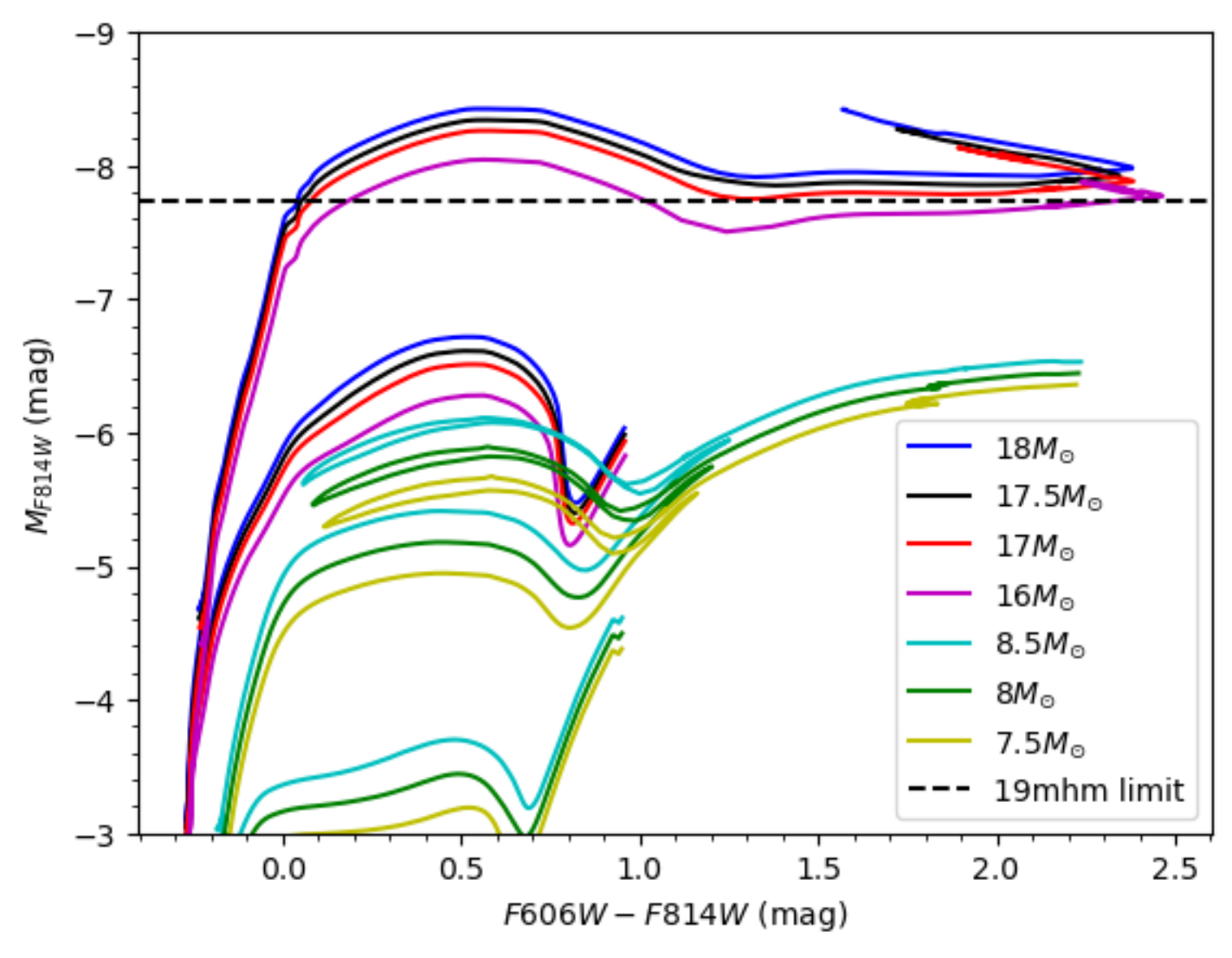}
\caption{Evolutionary tracks of various stellar masses. The absolute limit of SN\, 2019mhm is noted in blue.}
\label{fig:cmd}
\end{center}
\end{figure}

Using the deeper WFC3 F814W frame as a reference image for all sources detected in both the WFC3 and WFPC2 data, we obtained photometry of point sources in the individual flattened frames ({\tt flc/c0m}) using {\tt dolphot} \citep{dolphot}. We derived the upper limit on the presence of a point-like source in pre-explosion WFPC2 frames by injecting various sources with magnitudes from 19 to 26~mag into the individual flat frames at the location of SN\, 2019mhm. Based on the non-detection of a source in the WFPC2 frames, we derived an upper limit on the presence of a counterpart by analyzing sources detected at $\geq$3$\sigma$ significance within 20$^{\prime\prime}$ of the location of SN\,2019mhm. This estimate implies that in the pre-explosion image, any counterpart to SN\,2019mhm must have had $m_{\rm F814W} > 24.53$~mag (AB). We show the evolutionary tracks of stars with various masses in \autoref{fig:cmd}. In the post-explosion image (roughly 1.6~yr after discovery), we derive $m_{\rm F814W}=23.28\pm0.04$~mag (AB) for SN\,2019mhm itself (\autoref{tab:cand-photometry} and \autoref{fig:lc}).

Using the limits on progenitor flux along with the distance modulus, Milky Way extinction and extinction in the host galaxy, we considered giant star tracks and the in-band luminosities of these stars right before explosion. We compared our limits to the MESA Isochrones and Stellar Tracks (MIST) models \citep{Choi17}, and we found that our F814W explosion limit rules out all terminal RSGs with ZAMS masses $>$17.5~$M_{\odot}$. 

\begin{figure}
\begin{center}
\includegraphics[width=0.49\textwidth]{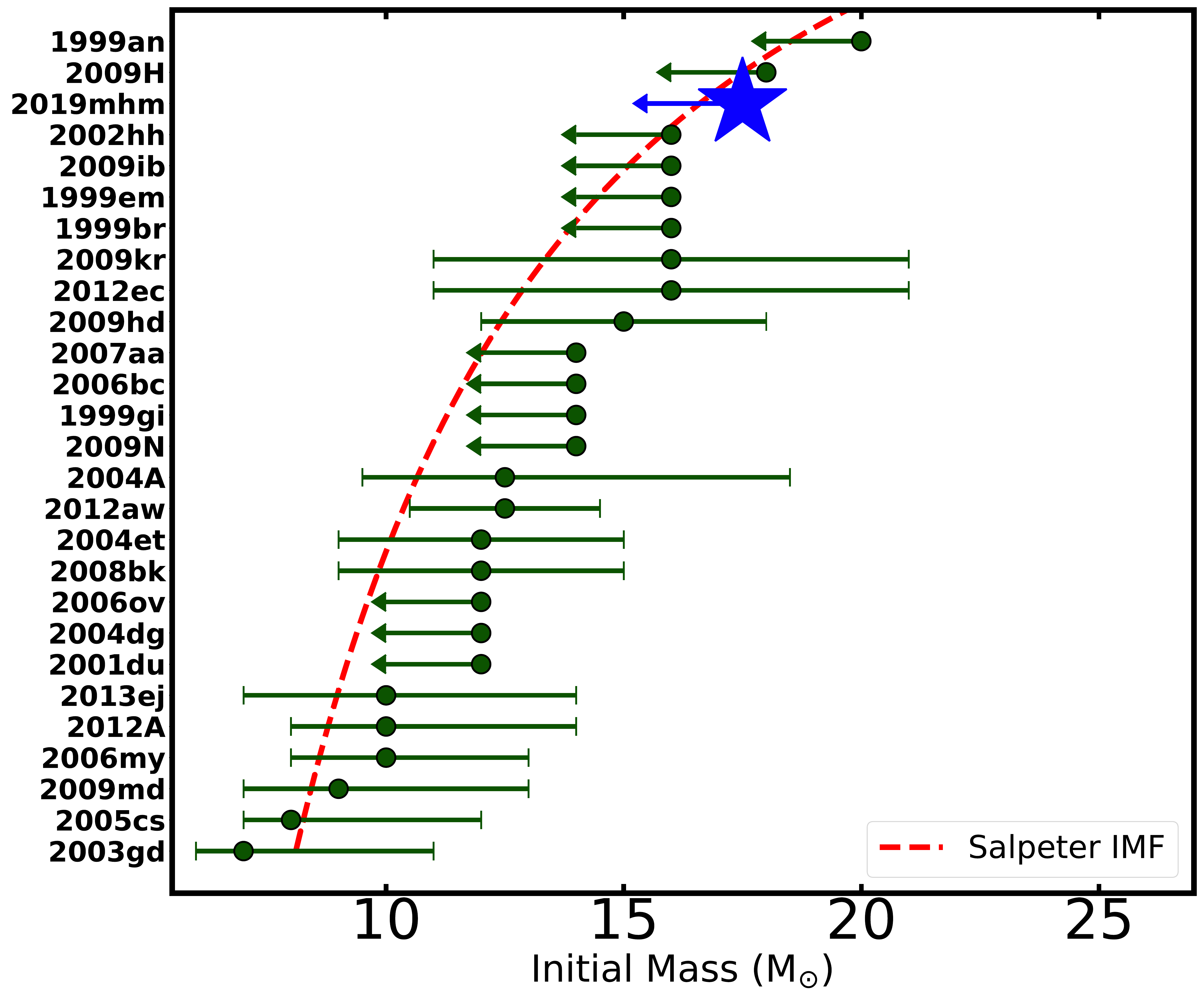}
\caption{Masses of RSG progenitor stars in the literature, with a Salpeter IMF truncated at 8 and 24~$M_{\odot}$.} 
\label{fig:IMF}
\end{center}
\end{figure}

The limit we derived is comparable to the upper limits for RSG progenitors presented in \cite{Smartt09, Smartt15}, but more recent analyses suggest that the mass limit is actually higher than 18~$M_{\odot}$. \cite{Davies18a} argue that systematic error leads to a significant underestimation of progenitor star masses, and present a cutoff at 25~$M_\odot$, with an upper limit of $<33~M_\odot$ at 2$\sigma$ confidence. \cite{Davies20} suggest that the lower mass limit is 6-8~$M_\odot$, and the upper mass limit between 18-20~$M_\odot$. It should be noted that as a whole, the sample size for detected progenitor stars is rather small, making the analysis of SNe like SN\,2019mhm important to help build up statistics. We present the masses of RSG stars in the literature in \autoref{fig:IMF}.

\section{Discussion \& Conclusions}
\label{sec:disc} 

SN\,2019mhm can be added to the current sample of direct detections and upper limits  used to analyze the Type II SN progenitor mass function, currently at 26 SNe \citep{Smartt15}. Based on the methods described above, we derived a ZAMS mass limit of 17.5~$M_{\odot}$. This further constrains the mass threshold at which Type II-P SNe can explode from RSGs as likely being below 18~$M_{\odot}$ assuming a Salpeter IMF. Indeed, \citet{Davies18a} argue that the ``upper mass'' threshold for the RSG counterparts to SNe\,II may not be statistically significant with an upper mass limit that is consistent with the maximum mass RSG at the 3$\sigma$ level \citep[i.e., at the Humphreys-Davidson limit of $ \approx$30~$M_{\odot}$;][]{Humphreys79}. 

Following methods in \citet{Davies20}, we analyzed the Type II progenitor star photometry presented in that paper to evaluate the effect of adding SN\,2019mhm on that analysis. In summary, this analysis takes the progenitor star photometry, extinction, distance, and bolometric correction for each detected source and upper limit \citep[24 total in][]{Davies20} and estimates the most likely observed progenitor star luminosity distribution by varying each of these quantities within their reported uncertainties. It then compares this distribution to an empirical luminosity power-law distribution with a power-law index ($\Gamma$), lower luminosity cutoff ($L_{\rm low}$) and upper luminosity cutoff ($L_{\rm high}$). This analysis removes systematic biases in translating from RSG luminosity to initial mass and can be compared directly to the observed luminosities of RSGs in the Milky Way and other galaxies.

Following this analysis, we derive a lower and upper luminosity cutoffs of $\log(L_{\rm low}/L_{\odot})=4.43\substack{+0.09\\-0.10}$ and $\log(L_{\rm high}/L_{\odot})=5.23\substack{+0.14\\-0.10}$, with nominally smaller uncertainties than in \citet{Davies20} but consistent with their analysis. When we added our limits on SN\,2019mhm and re-ran this analysis, we found these values changed to $\log(L_{\rm low}/L_{\odot})=4.43\substack{+0.07\\-0.10}$ and $\log(L_{\rm high}/L_{\odot})=5.21\substack{+0.12\\-0.10}$, in agreement with the previous analysis although with a moderately lower $L_{\rm high}$ value and smaller uncertainties. 

Analyzing every nearby SN with pre-explosion imaging is essential to building a large, unbiased population of progenitor systems and understand the causes of the RSG problem \citep{Smartt15}. Another possible solution to the RSG problem is that dust shells surrounding these stars obscure them, and as a result their luminosities and masses are severely underestimated. However, it was shown that the luminosities of these progenitors in the X-ray luminosities were not high enough to prove this was the case \citep{Dwarkadas14}. The conclusion that dust obscuration significantly lowers the inferred luminosities of these red supergiants assumes that the circumstellar material is extended around the progenitor star, whereas some Type II progenitors are known to have compact circumstellar shells \citep{Kilpatrick18}.
\acknowledgments
\section{Acknowledgements}
We thank Ben Davies for providing progenitor luminosity code and helpful comments on our analysis. We also thank Jorge Anais Vilchez, Abdo Campillay, Yilin Kong-Riveros, Nahir Muñoz-Elgueta and Natalie Ulloa for observations on the Swope Telescope at Las Campanas Observatory, Chile. We thank Tom Holoien and Subo Dong for providing raw duPont data. 
This material is based upon work supported by the National Science Foundation under grant No.1757792,a Research Experiences for Undergraduates (REU) grant awarded to CIERA at Northwestern University. Any opinions, findings, and conclusions or recommendations expressed in this material are those of the author(s) and do not necessarily reflect the views of the National Science Foundation.

\facilities{{\it HST} (WFPC2, WFC3), Swope (Direct)}

\software{{\tt dolphot} \citep{dolphot},
          {\tt DoPhot} \citep{Schechter93},
          {\tt hst123} \citep{Kilpatrick21},
          {\tt photpipe} \citep{Rest05}
          {\tt hotpants} \citep{hotpants}
          {\tt TweakReg} \citep{tweakreg}
          {\tt emcee} \citep{MCMC}
          {\tt BANZAI} \citep{BANZAI}}

\startlongtable
\begin{deluxetable}
{ccccc}
\tabletypesize{\scriptsize}
\tablecaption{Photometry of SN\,2019mhm\label{tab:cand-photometry}}
\tablewidth{0pt}
\tablehead{
\colhead{MJD} &
\colhead{Filter} &
\colhead{$m$} &
\colhead{$\sigma_{m}$} &
\colhead{Source}\\
&
&
(mag) &
(mag) &
}
\startdata
58698.0523 & $B$& $>$17.151 & -- & Swope \\
58699.1103 & $B$ & 17.195 & 0.015 & Swope \\
58700.1585 & $B$ & 17.203 & 0.014 & Swope \\
58701.0631 & $B$ & 17.248 & 0.017 & Swope \\
58702.0586 & $B$ & 17.244 & 0.019 & Swope \\
58703.0934 & $B$ & 17.296 & 0.027 & Swope \\
58704.1889 & $B$ & 17.317 & 0.021 & Swope \\
58711.0576 & $B$ & 17.647 & 0.044 & Swope \\
58712.0317 & $B$ & 17.686 & 0.029 & Swope \\
58713.0139 & $B$ & 17.742 & 0.020 & Swope \\
58714.0861 & $B$ & 17.757 & 0.028 & Swope \\
58716.0392 & $B$ & 17.820 & 0.017 & Swope \\
58718.0530 & $B$ & 17.920 & 0.017 & Swope \\
58725.0228 & $B$ & 18.070 & 0.017 & Swope \\
58745.0505 & $B$ & 18.356 & 0.026 & Swope \\
58748.0183 & $B$ & 18.424 & 0.019 & Swope \\
58753.9984 & $B$ & 18.546 & 0.021 & Swope \\
58760.0564 & $B$ & 18.753 & 0.104 & Swope \\
58775.0350 & $B$ & 18.918 & 0.031 & Swope \\
58784.0457 & $B$ & 19.157 & 0.027 & Swope \\
58805.0251 & $B$ & 19.936 & 0.059 & Swope \\
58698.0512 & $V$ & 16.881 & 0.017 & Swope \\
58699.1082 & $V$ & 16.866 & 0.018 & Swope \\
58699.1092 & $V$ & 16.874 & 0.017 & Swope \\
58700.1573 & $V$ & 16.899 & 0.016 & Swope \\
58701.0620 & $V$ & 16.896 & 0.018 & Swope \\
58702.0575 & $V$ & 16.845 & 0.023 & Swope \\
58703.0924 & $V$ & 16.846 & 0.025 & Swope \\
58704.1877 & $V$ & 16.832 & 0.020 & Swope \\
58711.0564 & $V$ & 16.925 & 0.037 & Swope \\
58712.0302 & $V$ & 16.939 & 0.023 & Swope \\
58713.0127 & $V$ & 16.981 & 0.019 & Swope \\
58714.0848 & $V$ & 16.989 & 0.030 & Swope \\
58716.0361 & $V$ & 17.024 & 0.015 & Swope \\
58718.0519 & $V$ & 17.029 & 0.017 & Swope \\
58725.0214 & $V$ & 17.087 & 0.015 & Swope \\
58745.0480 & $V$ & 17.148 & 0.019 & Swope \\
58748.0158 & $V$ & 17.169 & 0.016 & Swope \\
58753.9963 & $V$ & 17.249 & 0.016 & Swope \\
58760.0537 & $V$ & 17.269 & 0.088 & Swope \\
58775.0328 & $V$ & 17.335 & 0.024 & Swope \\
58784.0435 & $V$ & 17.558 & 0.019 & Swope \\
58805.0224 & $V$ & 18.472 & 0.032 & Swope \\
58698.0449 & $g$ & 16.869 & 0.018 & Swope \\
58699.1019 & $g$ & 16.932 & 0.016 & Swope \\
58700.1502 & $g$ & 16.933 & 0.014 & Swope \\
58701.0563 & $g$ & 16.954 & 0.016 & Swope \\
58702.0518 & $g$ & 17.024 & 0.021 & Swope \\
58703.0864 & $g$ & 17.159 & 0.051 & Swope \\
58704.1860 & $g$ & 17.165 & 0.022 & Swope \\
58711.0547 & $g$ & 17.292 & 0.052 & Swope \\
58712.0233 & $g$ & 17.243 & 0.025 & Swope \\
58713.0111 & $g$ & 17.234 & 0.018 & Swope \\
58714.0780 & $g$ & 17.280 & 0.025 & Swope \\
58716.0275 & $g$ & 17.334 & 0.014 & Swope \\
58718.0453 & $g$ & 17.378 & 0.016 & Swope \\
58725.0128 & $g$ & 17.495 & 0.015 & Swope \\
58737.2121 & $g$ & 17.575 & 0.020 & Swope \\
58745.0377 & $g$ & 17.677 & 0.021 & Swope \\
58748.0039 & $g$ & 17.745 & 0.015 & Swope \\
58760.0507 & $g$ & 17.707 & 0.137 & Swope \\
58761.0173 & $g$ & 17.787 & 0.024 & Swope \\
58775.0306 & $g$ & 17.993 & 0.027 & Swope \\
58784.0408 & $g$ & 18.191 & 0.019 & Swope \\
58791.0157 & $g$ & 18.355 & 0.042 & Swope \\
58805.0197 & $g$ & 19.122 & 0.039 & Swope \\
58815.0265 & $g$ & 19.982 & 0.049 & Swope \\
58910.3514 & $g$ & 20.738 & 0.090 & Swope \\
58923.4080 & $g$ & 20.927 & 0.165 & Swope \\
58698.0439 & $i$ & 16.783 & 0.018 & Swope \\
58699.1003 & $i$ & 16.779 & 0.017 & Swope \\
58700.1489 & $i$ & 16.782 & 0.016 & Swope \\
58701.0553 & $i$ & 16.739 & 0.017 & Swope \\
58702.0508 & $i$ & 16.729 & 0.020 & Swope \\
58703.0854 & $i$ & 16.708 & 0.035 & Swope \\
58704.1850 & $i$ & 16.686 & 0.019 & Swope \\
58711.0535 & $i$ & 16.676 & 0.034 & Swope \\
58712.0211 & $i$ & 16.687 & 0.022 & Swope \\
58713.0098 & $i$ & 16.665 & 0.018 & Swope \\
58714.0769 & $i$ & 16.670 & 0.018 & Swope \\
58716.0257 & $i$ & 16.681 & 0.015 & Swope \\
58718.0442 & $i$ & 16.654 & 0.015 & Swope \\
58725.0115 & $i$ & 16.644 & 0.015 & Swope \\
58737.2043 & $i$ & 16.603 & 0.014 & Swope \\
58745.0352 & $i$ & 16.611 & 0.016 & Swope \\
58748.0013 & $i$ & 16.616 & 0.014 & Swope \\
58753.9862 & $i$ & 16.617 & 0.017 & Swope \\
58760.0493 & $i$ & 16.616 & 0.109 & Swope \\
58775.0284 & $i$ & 16.678 & 0.017 & Swope \\
58784.0385 & $i$ & 16.789 & 0.017 & Swope \\
58791.0141 & $i$ & 16.883 & 0.028 & Swope \\
58805.0178 & $i$ & 17.464 & 0.024 & Swope \\
58815.0216 & $i$ & 18.184 & 0.027 & Swope \\
58910.3439 & $i$ & 19.481 & 0.051 & Swope \\
58923.3903 & $i$ & 19.567 & 0.111 & Swope \\
58923.4006 & $i$ & 19.618 & 0.050 & Swope \\
58698.0429 & $r$ & 16.769 & 0.016 & Swope \\
58699.0991 & $r$ & 16.788 & 0.017 & Swope \\
58700.1475 & $r$ & 16.765 & 0.014 & Swope \\
58701.0543 & $r$ & 16.752 & 0.016 & Swope \\
58702.0498 & $r$ & 16.760 & 0.020 & Swope \\
58703.0844 & $r$ & 16.726 & 0.023 & Swope \\
58704.1840 & $r$ & 16.693 & 0.018 & Swope \\
58711.0525 & $r$ & 16.736 & 0.030 & Swope \\
58712.0199 & $r$ & 16.759 & 0.018 & Swope \\
58713.0086 & $r$ & 16.781 & 0.018 & Swope \\
58714.0747 & $r$ & 16.755 & 0.018 & Swope \\
58716.0237 & $r$ & 16.800 & 0.014 & Swope \\
58718.0432 & $r$ & 16.805 & 0.014 & Swope \\
58725.0093 & $r$ & 16.821 & 0.014 & Swope \\
58737.1967 & $r$ & 16.822 & 0.014 & Swope \\
58745.0327 & $r$ & 16.857 & 0.016 & Swope \\
58747.9988 & $r$ & 16.845 & 0.013 & Swope \\
58753.9848 & $r$ & 16.876 & 0.015 & Swope \\
58760.0480 & $r$ & 16.830 & 0.048 & Swope \\
58761.0138 & $r$ & 16.831 & 0.055 & Swope \\
58775.0263 & $r$ & 16.946 & 0.017 & Swope \\
58784.0355 & $r$ & 17.043 & 0.015 & Swope \\
58791.0124 & $r$ & 17.162 & 0.023 & Swope \\
58805.0157 & $r$ & 17.791 & 0.021 & Swope \\
58814.0250 & $r$ & 18.388 & 0.023 & Swope \\
58815.0165 & $r$ & 18.472 & 0.028 & Swope \\
58904.3849 & $r$ & 19.675 & 0.073 & Swope \\
58904.3857 & $r$ & 19.748 & 0.029 & Swope \\
58910.3363 & $r$ & 19.759 & 0.033 & Swope \\
58923.3827 & $r$ & 19.845 & 0.044 & Swope \\
58923.3931 & $r$ & 19.853 & 0.039 & Swope \\
58698.0459 & $u$ & 17.799 & 0.018 & Swope \\
58699.1031 & $u$ & 17.865 & 0.017 & Swope \\
58700.1515 & $u$ & 17.948 & 0.016 & Swope \\
58701.0574 & $u$ & 18.025 & 0.019 & Swope \\
58702.0528 & $u$ & 18.135 & 0.025 & Swope \\
58703.0875 & $u$ & 18.266 & 0.037 & Swope \\
58712.0248 & $u$ & 19.705 & 0.108 & Swope \\
58714.0795 & $u$ & 19.774 & 0.085 & Swope \\
58716.0295 & $u$ & 19.905 & 0.042 & Swope \\
58718.0468 & $u$ & 20.204 & 0.052 & Swope \\
58725.0141 & $u$ & 20.666 & 0.071 & Swope \\
58745.0402 & $u$ & 21.357 & 0.181 & Swope \\
58748.0069 & $u$ & 21.463 & 0.138 & Swope \\
58753.9889 & $u$ & 21.638 & 0.181 & Swope \\
58697.8094 & $B$ & 17.114 & 0.028 & LCO \\
58698.6828 & $B$ & 17.115 & 0.026 & LCO \\
58698.6845 & $B$ & 17.138 & 0.027 & LCO \\
58699.7424 & $B$ & 17.111 & 0.027 & LCO \\
58701.2949 & $B$ & 17.207 & 0.032 & LCO \\
58713.8128 & $B$ & 17.709 & 0.030 & LCO \\
58719.1947 & $B$ & 17.956 & 0.031 & LCO \\
58724.1995 & $B$ & 18.105 & 0.031 & LCO \\
58757.8405 & $B$ & 18.743 & 0.047 & LCO \\
58760.7363 & $B$ & 18.704 & 0.046 & LCO \\
58772.7468 & $B$ & 18.867 & 0.050 & LCO \\
58791.0465 & $B$ & 19.269 & 0.100 & LCO \\
58697.8117 & $V$ & 16.929 & 0.025 & LCO \\
58698.6864 & $V$ & 16.896 & 0.026 & LCO \\
58698.6877 & $V$ & 16.910 & 0.026 & LCO \\
58699.7448 & $V$ & 16.889 & 0.025 & LCO \\
58701.2972 & $V$ & 16.873 & 0.024 & LCO \\
58713.8147 & $V$ & 16.967 & 0.025 & LCO \\
58719.1966 & $V$ & 17.032 & 0.024 & LCO \\
58724.2014 & $V$ & 17.038 & 0.020 & LCO \\
58757.8436 & $V$ & 17.226 & 0.023 & LCO \\
58760.7394 & $V$ & 17.170 & 0.023 & LCO \\
58764.7429 & $V$ & 17.220 & 0.023 & LCO \\
58768.7399 & $V$ & 17.276 & 0.027 & LCO \\
58772.7499 & $V$ & 17.327 & 0.026 & LCO \\
58698.6893 & $g$ & 16.967 & 0.022 & LCO \\
58698.6910 & $g$ & 16.952 & 0.021 & LCO \\
58713.8161 & $g$ & 17.243 & 0.025 & LCO \\
58719.1980 & $g$ & 17.391 & 0.025 & LCO \\
58724.2028 & $g$ & 17.417 & 0.024 & LCO \\
58757.8453 & $g$ & 17.808 & 0.027 & LCO \\
58760.7410 & $g$ & 17.826 & 0.026 & LCO \\
58764.7446 & $g$ & 17.850 & 0.028 & LCO \\
58772.7515 & $g$ & 17.983 & 0.028 & LCO \\
58791.0491 & $g$ & 18.357 & 0.043 & LCO \\
58801.0107 & $g$ & 18.950 & 0.115 & LCO \\
58803.7668 & $g$ & 19.094 & 0.126 & LCO \\
58808.0161 & $g$ & 19.263 & 0.272 & LCO \\
58697.8082 & $i$ & 16.782 & 0.037 & LCO \\
58698.6957 & $i$ & 16.778 & 0.046 & LCO \\
58698.6971 & $i$ & 16.770 & 0.043 & LCO \\
58699.7413 & $i$ & 16.718 & 0.030 & LCO \\
58701.2937 & $i$ & 16.708 & 0.031 & LCO \\
58709.9815 & $i$ & 16.673 & 0.032 & LCO \\
58713.8189 & $i$ & 16.625 & 0.030 & LCO \\
58719.2008 & $i$ & 16.643 & 0.025 & LCO \\
58724.2056 & $i$ & 16.596 & 0.024 & LCO \\
58757.8487 & $i$ & 16.573 & 0.027 & LCO \\
58760.7445 & $i$ & 16.539 & 0.025 & LCO \\
58764.7480 & $i$ & 16.577 & 0.025 & LCO \\
58768.7450 & $i$ & 16.614 & 0.027 & LCO \\
58772.7550 & $i$ & 16.625 & 0.031 & LCO \\
58791.0523 & $i$ & 16.888 & 0.030 & LCO \\
58801.0130 & $i$ & 17.142 & 0.039 & LCO \\
58803.7691 & $i$ & 17.459 & 0.086 & LCO \\
58808.0189 & $i$ & 17.761 & 0.049 & LCO \\
58910.3803 & $i$ & 19.541 & 0.173 & LCO \\
58697.8105 & $r$ & 16.841 & 0.023 & LCO \\
58698.6928 & $r$ & 16.830 & 0.023 & LCO \\
58698.6942 & $r$ & 16.811 & 0.022 & LCO \\
58699.7436 & $r$ & 16.755 & 0.023 & LCO \\
58701.2960 & $r$ & 16.733 & 0.021 & LCO \\
58713.8176 & $r$ & 16.759 & 0.021 & LCO \\
58719.1995 & $r$ & 16.835 & 0.021 & LCO \\
58724.2043 & $r$ & 16.803 & 0.018 & LCO \\
58757.8473 & $r$ & 16.849 & 0.022 & LCO \\
58760.7430 & $r$ & 16.830 & 0.019 & LCO \\
58764.7466 & $r$ & 16.880 & 0.021 & LCO \\
58768.7436 & $r$ & 16.866 & 0.022 & LCO \\
58772.7535 & $r$ & 16.901 & 0.021 & LCO \\
58791.0508 & $r$ & 17.165 & 0.025 & LCO \\
58801.0119 & $r$ & 17.536 & 0.035 & LCO \\
58803.7680 & $r$ & 17.760 & 0.048 & LCO \\
58808.0176 & $r$ & 18.157 & 0.042 & LCO \\
58910.3762 & $r$ & 19.797 & 0.087 & LCO \\
59309.8452 & F814W & 23.282 & 0.039 & {\it HST} \\
59309.8515 & F300X & 27.202 & 0.368 & {\it HST} \\
\enddata
\tablecomments{Photometry of SN\,2019mhm from {\it HST}, the Swope 1m telescope, and Las Cumbres Observatory (LCO) 1m telescopes. All photometry is on the AB mag system.}
\end{deluxetable}
\bibliography{references}

\end{document}